\documentstyle[twocolumn,prd,aps,epsf,psfrag,floats]{revtex}
\begin{document}
\draft
\preprint{UWThPh -- 1996 -- 38}
\title{Initial data for general relativity\\
       containing a marginally outer trapped torus}
\author{S. Husa}
\address{Institut f\"ur Theoretische Physik  \\
         Universit\"at Wien \\
         Boltzmanngasse 5, A-1090 Wien, Austria \\
         e-mail: shusa@merlin.pap.univie.ac.at}
\date{June 14, 1996}
\maketitle
\begin{abstract}
Asymptotically flat, time-symmetric, axially symmetric and conformally flat
initial data for vacuum general relativity are studied numerically on $R^3$
with the interior of a standard torus cut out. By the choice of boundary
condition the torus is marginally outer trapped, and thus a surface of minimal
area. Apart from pure scaling the standard tori are parameterized by a radius
$a\in [0,1]$, where $a=0$ corresponds to the limit where the boundary torus
degenerates to a circle and $a=1$ to a torus that touches the axis of symmetry.
Noting that these tori are the orbits of a $U(1)\times U(1)$ conformal isometry
allows for a simple scheme to solve the constraint, involving numerical
solution of only ordinary differential equations. Regular solutions can be
constructed for $a\in\, ]a_{0}\sim 0.14405,1[\,$. The tori are unstable minimal
surfaces (i.e. only saddle points of the area functional) and thus can not be
apparent horizons, but they are always surrounded by an apparent horizon of
spherical topology, which is analyzed in the context of the hoop conjecture
and isoperimetric inequality for black holes.
\end{abstract}
\pacs{PACS numbers: 4.20.Ex, 2.40.Ky, 4.25.Dm}
%
%
\section{Introduction}

The aim of this paper is to construct a class of asymptotically flat solutions
to the vacuum constraint equations of general relativity that are somewhat
unusual, namely they contain a marginally outer trapped torus.
In the case of time symmetry discussed here, a marginally outer trapped
surface has vanishing mean curvature and is a surface of minimal area
(see Ref. \cite{galloway} and references cited therein).
The minimal tori are all unstable, that is, they are only
saddle points of the area functional. In a recent discussion of
minimal area tori and their connection with topological censorship
Galloway \cite{galloway} proved, that an unstable minimal torus locally can
be deformed into an outer trapped surface. The outer boundary of the region
containing trapped surfaces -- the apparent horizon -- is a stable marginally
outer trapped surface of spherical topology (see Sec. \ref{ah}).

Motivation for this work comes from two sources: while the question of the
existence of a regular solution to the constraints is of physical and
mathematical significance in itself, analysis of given initial data with
respect to the existence and properties of minimal surfaces and apparent
horizons has received some interest in the discussion on cosmic
censorship.

This work deals with the simplest situation where a toroidal minimal surface
can be present, namely that of time-symmetry, axial symmetry and conformal
flatness.
The assumptions of time-symmetry and conformal flatness allow to use a
conformal symmetry to turn the constraint from a partial differential equation
(PDE) to a set of uncoupled ordinary differential equations (ODEs).
Generalizations of the scheme that relax the restriction to conformal flatness
and axial symmetry will be discussed below in Sec. \ref{discussion}.

The physical data are given by an asymptotically flat metric $\tilde g$
defined on a manifold $\tilde M$ that is taken to be $R^{3}$ with a region
inside a torus $\partial\tilde M$ cut out.
The physical metric $\tilde g$ is required to satisfy the vacuum Hamiltonian
constraint of vanishing scalar curvature $R_{\tilde g}$ on $\tilde M$
plus the boundary condition of vanishing mean curvature $\tilde p/2$
($\tilde p$ is the trace of the extrinsic curvature) at
$\partial \tilde M$, making it a minimal surface:
\begin{eqnarray}
\label{problem_1}
    R_{\tilde g} & = & 0 \qquad\mbox{on $\tilde M$}, \\
\label{problem_2}
    \tilde p     & = & 0 \qquad\mbox{on $\partial\tilde M$}.
\end{eqnarray}

The standard way to solve these equations is the conformal approach
\cite{conf_approach}. Restricting to conformal flatness we
select the flat metric $\delta={dr}^{2}+{dz}^{2}+r^{2}{d\varphi}^{2}$
on $\tilde M$ as our base metric and
construct the physical metric as $\tilde g=\psi^4\, \delta$,
which reduces Eq. (\ref{problem_1}) to an
elliptic PDE for the conformal factor $\psi$.

For notational convenience we define the linear differential operators
\begin{equation}\label{def_L}
L_{g}:= -\bigtriangleup_{g} + \frac{1}{8}R_{g},
\end{equation}
the conformal Laplacian of a metric $g$, and
\begin{equation}\label{def_B}
B_{g}:= \frac{\partial}{\partial n} + \frac{1}{4}p,
\end{equation}
where ${\partial}/{\partial n}$ is the normal derivative and $p$ is the trace
of the extrinsic curvature (or twice the mean curvature) induced by the
3-metric $g$ on a hypersurface.
Using the definitions of $L_{g}$ and $B_{g}$ the constraint equation
(\ref{problem_1}) with boundary condition (\ref{problem_2}) turns into the
problem of finding a conformal factor $\psi$, such that
\begin{eqnarray}\label{problem_conformal_01}
 L_{\delta}\psi \equiv -\bigtriangleup_{\delta}\psi & = & 0\qquad\mbox{on
$\tilde M$},\\
\label{problem_conformal_02}
 B_{\delta}\psi & = & 0\qquad\mbox{on $\partial \tilde M$},\\
\label{problem_conformal_03}
\psi&>&0,\qquad \psi\rightarrow 1\quad\mbox{as}\quad
r^{2}+z^{2}\rightarrow\infty.
\end{eqnarray}

We are left with the freedom to fix the boundary surface,
taking it to be a standard torus in flat space, defined in cylindrical
coordinates as a solution of
\begin{equation}\label{def_bound}
z^2 + (r - A)^2 = a^2.
\end{equation}
Since a finite overall scaling of the coordinates bears no physical
significance, we can fix
$A=1$ for the boundary tori considered, which leaves us with the one
parameter family $0\leq a \leq 1$. The extremal case $a=0$ corresponds
to a circle of unit radius in the $z=0$ plane, the other extreme,
$a=1$, defines a torus which touches its axis of symmetry.

Using the method described below it is possible to continue the data smoothly
inside the minimal tori to some finite extent, assuming vacuum and conformal
flatness inside,
but in general this assumption will lead to a nodal surface, and thus to a
curvature singularity, if one proceeds too far. The obvious question whether
it is possible to get a regular interior {\em and} nonnegative
energy density of the matter is not dealt with here.

Regular solutions to the constraint equation (\ref{problem_conformal_01})
with boundary conditions (\ref{problem_conformal_02},
\ref{problem_conformal_03}) exist for $a\in]a_{0}\sim 0.14405,1[$ (all
numerical results are rounded to the last digit), where the parameter $a$
labels the boundary tori as defined above in Eq. (\ref{def_bound}). In the
limit $a\rightarrow a_{0}$ the mass blows up, and the Sobolev quotient (an
extension of the definition of the Yamabe number to manifolds with
boundary) vanishes. Also the apparent horizon moves outward to infinity and
becomes spherical in the limit. This behavior is familiar from the case when
the (usual) Yamabe number approaches zero \cite{beig+om}.
If $a$ is further decreased, the Sobolev quotient
becomes negative, the mass diverges to minus infinity as $a\nearrow a_{0}$,
and the conformal factor develops a nodal surface. In the other extremal
case,
when $a\rightarrow 1$, the mass stays finite, but a curvature singularity
develops at the center, where the toroidal minimal surface touches the axis
of symmetry. In this last case the solution can be given explicitly.

The toroidal minimal surfaces are all unstable, and part of their spectrum of
perturbations was computed.

The ADM mass can be
cast into an explicitly positive form in a way similar to Brill's result
\cite{brill}, additionally it can be bounded from below.

Note that the initial data discussed here are perfectly viable for
evolution, since an apparent horizon boundary condition \cite{ahbc}
can be applied at the topologically spherical apparent horizon that
surrounds the boundary torus. The minimal torus could
then be viewed as a model of a bulk of matter or gravitational radiation
that has enough energy to confine all information about it inside of the
horizon. This comment also holds in the case $a\rightarrow 1$, when a
curvature singularity appears at the center.

By using conformal symmetries, the PDE can be decomposed into an infinite
set of uncoupled linear ODEs that were solved numerically.
A convenient cutoff of the infinite series yields a highly accurate
solution for the constraint. The existence of solutions can be decided by
solving a single linear ODE, together with the high accuracy obtained for the
actual solution, this enables a secure grip on the two interesting limiting
cases.

The key technical input is the utilization of a $U(1)\times U(1)$ conformal
isometry that leaves the boundary torus invariant.
In order to benefit from the conformal isometry the conformal method is
reformulated for a compact base manifold $M$, taken to be the one-point
compactification of $\tilde M$, that is $S^3$ with the interior of a torus
cut out.
The conformal ansatz $\tilde g = G^4 g$, with $g$ the standard metric on
$S^3$, results in an elliptic PDE similar to (\ref{problem_conformal_01}),
\begin{equation}\label{problem_conformal_1}
L_{g}G = 4\pi c\,\delta_{\Lambda} \qquad \mbox{on $M$},
\end{equation}
but with a distributional source term: $\delta_{\Lambda}$ is the Dirac delta
distribution concentrated at the point $\Lambda$ on $M$, which corresponds to
asymptotic infinity of $\tilde M$, and $c$ is a positive constant that can be
chosen arbitrarily for overall scaling. $G$ is thus a Green function for
$c^{-1}L_{g}$ with boundary condition
\begin{equation}\label{problem_conformal_2}
B_{g}G = 0 \qquad \mbox{on $\partial M$}.
\end{equation}

The existence of a positive Green function $G$ is determined by the sign of
$\lambda_{1}$, the first eigenvalue of the conformal Laplacian with the
present boundary condition, or equivalently by the sign of the Sobolev
quotient $Q(g)$, a conformal invariant defined as
\begin{equation}\label{sobolevQ}
Q(g):=\inf_{\varphi}\frac{
\int_{M}(\vert\nabla\varphi\vert^{2} +
\frac{1}{8}R_{g}\varphi^{2}) d\,v
+ \frac{1}{4}\int_{\partial M} p \varphi^{2} d\,\sigma
}
{(\int_{M}\varphi^{6}d\,v)^{1/3}},
\end{equation}
which is a generalization of the Yamabe functional \cite{yamabe}
(defined for manifolds without boundary) to manifolds with boundary
\cite{escobar} and has the same sign as $\lambda_{1}$.

It can be shown in a manner equivalent to the no-boundary case dealt with in
\cite{lp}, that a positive $G$ exists iff $\lambda_{1}$ is positive.
Furthermore, when $G>0$, the positive mass theorem holds for
$\tilde g = G^4 g$ \cite{sy}.

The rest of this paper is organized as follows:
In the next section (\ref{comp_and_conf}) conformal compactification via
inverse stereographic projection from $R^3$ to $S^3$ and toroidal coordinates
thereon will be discussed.
In Sec. \ref{solution} the method to solve the constraint numerically
using toroidal coordinates will be developed, including numerical solution
for the first eigenvalue of the conformal Laplacian.
Section \ref{instability} deals with the instability of the minimal tori
by considering normal variations of the surfaces.
Apparent horizons are located and analyzed with respect to the
isoperimetric inequality for black holes \cite{penrose} and the
hoop conjecture \cite{thorne} in Sec. \ref{ah}. A discussion of the
results, with emphasis on the limiting cases, is given in section
\ref{discussion}.

\section{Compactification and Conformal Symmetry}\label{comp_and_conf}

For every choice of boundary torus $\Sigma_{\gamma}$, that is
a torus described by Eq. (\ref{def_bound})
with $A=1$ and parameterized by $\gamma:=\sqrt{1-a^{2}}\leq 1$,
we define a conformal rescaling from the flat metric
$\delta$ on $R^3$ to a metric $g(\gamma)$ defined on $S^3$ by
\begin{equation}\label{compactification}
g(\gamma)=\Omega(r,z;\gamma)^{2}\,\delta,
\qquad \Omega(r,z;\gamma)=\frac{2}{\gamma^2+r^2+z^2},
\end{equation}
which corresponds to an inverse stereographic projection.
The notation $f(\bar r,\bar z;\gamma)$ is chosen to denote dependence of a
function $f$ on both the spatial coordinates and the parameter $\gamma$,
frequently this will be shortened to writing $f(\gamma)$.
With the chosen normalization $g(\gamma)$ is the standard metric
on the 3-sphere of radius $1/\gamma$, which is conveniently written as
\begin{equation}\label{defmetric}
g(\gamma) = \frac{{d\bar r}^{2}}{1-\gamma^{2}\bar r^{2}} +
      (1-\gamma^{2}\bar r^{2}){d\bar z}^{2} + \bar r^{2}{d\varphi}^{2}
\end{equation}
in toroidal coordinates $(\bar r,\bar z,\varphi)$
taking ranges $0\leq\bar r\leq 1/\gamma$, $0\leq\gamma\bar z,\varphi\leq
2\pi$.
A discussion of the construction of this coordinate system on $S^3$ of unit
radius is given in \cite{tid}. The coordinate
$\bar r$ as a function on the $(r,z)$-half plane can be read off from Eqs.
(\ref{compactification}) and (\ref{defmetric}) yielding
\begin{displaymath}
\bar r(r,z;\gamma)=\Omega(r,z;\gamma)\, r.
\end{displaymath}
The relation between the coordinate $\bar z$ and the original cylindrical
coordinates is
\begin{displaymath}
\cos(\gamma\bar z) = \frac{1-\gamma^{2}\Omega(r,z;\gamma)}
{\sqrt{1 - \gamma^{2}\bar r(r,z;\gamma)^2}},
\qquad
\mbox{sign}(\bar z)=\mbox{sign}(z).
\end{displaymath}
The point $\Lambda$ on $S^{3}$, corresponding to asymptotic infinity of
$R^3$, is given by $\bar r=\bar z=0$.

The vector fields $(\partial/\partial\varphi)^{a}$, $(\partial /
\partial\bar z)^{a}$ form a
pair of commuting, orthogonal and hypersurface-orthogonal Killing vector
fields of $g(\gamma)$ spanning the surfaces of constant $\bar r$. For $\bar r
\neq 0,1/\gamma$ these are flat tori of constant mean curvature changing sign
at the so called Clifford torus for which $\bar r={1}/{\gamma\sqrt 2}$.
Going back to the flat metric and cylindrical coordinates,
a torus $\bar r=const.$ is mapped to a standard torus as defined by Eq.
(\ref{def_bound}) with parameters $A=1/\bar r$,
$a=\sqrt{1-\gamma^{2}\bar r^{2}}/\bar r$, in particular the torus $\bar r=1$
is mapped to the boundary torus $\Sigma_{\gamma}$.
The sets $\bar r=0$ (respectively $\bar r=1/\gamma$) are linked great circles
on $S^{3}$, corresponding to the $z$-axis
(respectively the circle $z=0,r=\gamma$)
after stereographic projection.

The totally geodesic surfaces $\bar z=const.$ ($\varphi=const.$) are metric
hemispheres intersecting at $\bar r=1/\gamma$ ($\bar r=0$) with
$\{\gamma\bar z=\alpha\}\cup\{\gamma\bar z=\alpha+\pi\}$
($\{\varphi=\beta\}\cup\{\varphi=\beta+\pi\}$) being
smoothly embedded $S^{2}$'s.
Under stereographic projection, the surfaces $\gamma\bar z=const.$
do not change topology except for $\{\gamma\bar z=0\}\cup\{\gamma\bar z=\pi\}$
which gets decompactified into
$\{z=0, r\geq \gamma\}\cup\{z=0, r\leq \gamma\}$, i.e.
the equatorial plane $z=0$. In flat space $\{\gamma\bar z=\alpha\}
\cup\{\gamma\bar z=\alpha+\pi\}$
is a sphere centered at $r=0$, $z=\gamma\cot{\gamma\bar z}$ with radius
$\sqrt{\gamma^{2}(\cot^{2}\gamma\bar z -1 )+2}$.

Note that for $\gamma=0$ the radius of $S^{3}$ becomes infinite, corresponding
to $R^{3}$, and the coordinates $\bar r$, $\bar z$ become inverted cylindrical
coordinates thereon:
\begin{displaymath}
\bar r=\frac{2r}{r^{2}+z^{2}},\qquad \bar z=\frac{2z}{r^{2}+z^{2}}.
\end{displaymath}
The tori $\bar r=const.$, corresponding to $A=a=1/\bar r$, all touch the
$z$-axis in this limiting case.

We have thus constructed a family of foliations of $R^3$ by standard tori,
with each element defined by selection of a boundary torus $\Sigma_{\gamma}$
that is contained in the foliation.
For every foliation labeled by $\gamma$ we defined in Eq.
(\ref{compactification}) a conformal rescaling with
compactifying factor $\Omega(r,z;\gamma)$. Since $\Omega$ is independent of
$\varphi$, $(\partial /\partial\varphi)^{a}$ is a Killing vector of $g$, but
$(\partial /\partial\bar z)^{a}$ is only a {\em conformal} Killing vector of
$g$.

The reason for the chosen scaling of $S^3$ as opposed to working on $S^3$
of fixed radius can bee seen when rescaling to the standard metric
$\hat g$ on $S^{3}$ of unit radius (with the notation of \cite{tid}):
\begin{displaymath}
\hat g := \gamma^{2}g(\gamma)=\frac{d\,\rho^{2}}{1-\rho^{2}}
 + (1-\rho^{2})d\,\chi^{2} + \rho^{2}d\,\varphi^{2},
\end{displaymath}
using rescaled toroidal coordinates $(\rho,\chi,\varphi)$
\begin{displaymath}
\rho(r,z;\gamma) := \gamma\,\bar r(r,z;\gamma),
\qquad\chi(r,z;\gamma) := \gamma\,\bar z.
\end{displaymath}
taking ranges $0\leq\rho\leq 1$, $0\leq\chi,\varphi\leq 2\pi$. The boundary
torus $\Sigma_{\gamma}$ that was defined by $\bar r=1$ before is now specified
by $\rho=\gamma$, so that the region of physical interest,
$0\leq\rho\leq\gamma$ shrinks to zero volume in the limit
$\gamma\rightarrow 0$. The problem of finding a conformal factor solving the
rescaled equation $L_{\hat g}\hat G=4\pi{\sqrt 2}\delta_{\Lambda}$ then becomes
ill-defined, the conformal factor $\hat G$ (and thus the mass
of ${\hat G}^{4}\hat g$) blows up for $\gamma\rightarrow 0$.

In order to formulate the constraint equation explicitly we need to write
down the operators $L_{g}$ and $B_{g}$ in our toroidal coordinate system.
The extrinsic curvature of a torus of constant $\bar r$
with respect to the metric $g_{ab}$ is (the semicolon denotes the covariant
derivative as usual)
\begin{equation}\label{pab}
p_{ab} = \bar r\sqrt{1-\gamma^{2}\bar r^{2}}\left(\varphi_{;a}
\varphi_{;b} -
\gamma^{2}\bar z_{;a}\bar z_{;b}\right).
\end{equation}
At $\partial M$ ($\bar r=1$) this yields
\begin{equation}\label{p}
p = p_{a}^{a} = \frac{1-2\gamma^{2}}{\sqrt{1-\gamma^{2}}},
\end{equation}
which vanishes at the Clifford torus given by $\gamma_{c}=1/\sqrt 2$, where
$\partial M$ is maximal. The extrinsic curvature $\tilde p_{ab}$ with respect
to the physical metric $\tilde g$ is
\begin{equation}\label{pab_physical}
\tilde p_{ab} = G^{2}\left( p_{ab} + 2 g_{ab} n^{c} G_{;c}
\right),
\end{equation}
where $n^{c}$ is the unit normal vector field on $\partial M$.
Note that since $p_{ab}$ has Lorentz signature, in particular it is not
proportional to $g$ (\ref{pab}), by the conformal rescaling formula for the
extrinsic curvature (\ref{pab_physical}) it is not possible to get
$\tilde p_{ab}=0$ by a conformal transformation. As a consequence the
minimal tori cannot be stable, since in that case they would have to be
totally geodesic ($\tilde p_{ab}=0$) -- (see \cite{galloway}).

Finally we can write down the operators $L_{\gamma}:=L_{g_{\gamma}}$ and
$B_{\gamma}:=B_{g_{\gamma}}$ -- simplifying their labels -- defined in
Eqs. (\ref{problem_conformal_1}, \ref{problem_conformal_2}) explicitly:
\begin{equation}\label{B}
B_{\gamma} = \sqrt{1-\gamma^{2}}\left(\frac{\partial}{\partial\bar r} +
   \frac{1-2\gamma^{2}}{4\sqrt{1-\gamma^{2}}}\right),
\end{equation}
\begin{eqnarray}\label{L}
L_{\gamma} = & - &(1-\gamma^{2}\bar r^{2})\frac{\partial^2}
            {\partial\bar r^{2}} - \frac{1-3\gamma^{2}\bar r^{2}}
            {\bar r}\frac{\partial}{\partial\bar r}  \nonumber  \\
  & - & \frac{1}{1-\gamma^{2}\bar r^{2}}\frac{\partial^{2}}
        {\partial\bar z^{2}}
    - \frac{1}{\bar r^{2}}\frac{\partial^{2}}{\partial\varphi^{2}}
    + \frac{3}{4}\gamma^{2}.
\end{eqnarray}

\section{Solution of the Constraint Equation}\label{solution}

\subsection{Formulation}

 From the choices made above our physical data are determined by the single
parameter $\gamma$, which labels the position of the boundary torus.
This defines the physical metric $\tilde g(\gamma)=G(\gamma)^{4} g(\gamma)$,
where $G(\gamma)$ satisfies the equations
\begin{eqnarray}\label{problem1}
L_{\gamma}G(\gamma) & = & 4\pi\sqrt{2}\, \delta_{\Lambda}
\qquad \mbox{on $M$}, \\ \label{problem2}
B_{\gamma}G(\gamma) & = & 0 \qquad \mbox{on $\partial M$}
\end{eqnarray}
with operators $L_{\gamma}$ and $B_{\gamma}$ given by the
expressions (\ref{B},\ref{L}), the boundary torus
is located at the fixed coordinate value $\bar r=1$ and the scale factor is
chosen as $c=\sqrt{2}$. The rescaling to flat space defined in Eq.
(\ref{compactification}) corresponds to a Green function $G_{0}$,
\begin{displaymath}
\qquad G_{0}(\gamma) = \Omega(\gamma)^{-1/2}=
\gamma\left(1-\sqrt{1-\gamma^{2}\bar r^{2}}\cos{\gamma\bar z}\right)^{-1/2},
\end{displaymath}
where $G_{0}(\gamma)$ satisfies Eq. (\ref{problem1})
but with the boundary condition (\ref{problem2})
replaced by regularity on $S^{3}\backslash\{\Lambda\}$.
The asymptotic behavior of $G$ near $\Lambda$ is known \cite{lp} to be
\begin{equation}\label{G_near_Lambda}
G = G_{0} + \frac{m}{2\sqrt{2}} + O(G_{0}^{-1}),
\end{equation}
where $m$ is the ADM-mass of the physical metric
$\tilde g$. By the positive mass theorem \cite{pos_mass} $m$ is positive
provided $\lambda_{1}$ is positive.

The relation to the original formulation of the constraint problem in Eqs.
(\ref{problem_conformal_01}-\ref{problem_conformal_03}) is as follows:
\begin{displaymath}
\tilde g=G^{4}g=\psi^{4}\delta=\psi^{4}\Omega^{-2}g\Rightarrow
\psi=\Omega^{-1/2}G,
\end{displaymath}
\begin{displaymath}
\psi = 1 + \frac{m}{2\sqrt{r^{2}+z^{2}}} + O(\Omega).
\end{displaymath}

The limiting cases $\gamma=0,1$ can be considered as singular situations, and
the question arises how this will effect the physical metric
$\tilde g(\gamma)$, in particular its existence.
For $\gamma<\gamma_{c}$ we have $p$ positive (see Eq. (\ref{p})) whence is the
Sobolev quotient $Q(g)$ defined in Eq. (\ref{sobolevQ}) -- a positive Green
function thus exists and the mass of the physical metric $\tilde g(\gamma)$ is
positive.
On the other hand $p$ is negative for $\gamma>\gamma_{c}$, so that a change of
sign becomes possible for $Q(g)$ at some $\gamma$,
$\gamma_{c}<\gamma<1$. This indeed happens at the numerically calculated value
$\gamma_{0}=0.98957$, corresponding to $a=0.14405$. As one expects, this case
is  analogous to the Yamabe number
going to zero in the usually considered case of manifolds without boundary.
The mass diverges to plus infinity as $\gamma$ approaches $\gamma_{0}$ from
below, and to minus infinity when taking the limit from above.

For $\gamma\rightarrow 0$ the metric $g$ becomes the flat metric on the
infinite cylinder $0\leq\bar r\leq1$, $-\infty<\bar z<\infty$,
$0\leq\varphi\leq 2\pi$:
\begin{displaymath}
g(0) = {d\bar r}^{2} + {d\bar z}^{2} + \bar r^{2}{d\varphi}^{2}.
\end{displaymath}
The point $\Lambda$, representing physical infinity, is at $\bar r=\bar z=0$,
the unphysical infinity, $\bar z=\pm\infty$ corresponds to the physical
center $r=z=0$.
The constraint Eq. (\ref{problem_conformal_1}) reduces to the cylindrically
symmetric Poisson equation for a point source
\begin{equation}\label{problem_g0_1}
\lefteqn{} L_{0}G(0) =
 -\left(\frac{\partial^{2}}{\partial \bar r^{2}} + \frac{1}{\bar r}
    \frac{\partial}{\partial\bar r}
 + \frac{\partial^{2}}{\partial\bar z^{2}}\right)G(0)
= 4\pi\sqrt{2}\delta_{\Lambda},
\end{equation}
with boundary condition
\begin{equation}\label{problem_g0_2}
\left(\frac{\partial }{\partial\bar r}
+\frac{1}{4}\right)G(0)\vert_{\bar r=1}=0.
\end{equation}
The explicit limiting solution is derived in Appendix \ref{expl_sol} as
\begin{eqnarray*}
\lefteqn{\frac{1}{\sqrt{2}}G(\bar r,\bar z;0) =
\frac{1}{\sqrt{\bar r^{2}+\bar z^{2}}} } \nonumber\\
& &\qquad +\frac{2}{\pi}\int_{0}^{\infty}\cos({\omega\bar z})
\frac{4\omega\left(K_{1}(\omega)-K_{0}(\omega)\right)}
{4\omega I_{1}(\omega) + I_{0}(\omega)} I_{0}(\omega\bar r)\,d\omega.
\end{eqnarray*}
As is discussed in Appendix \ref{expl_sol}, $G(\bar r,\bar z;0)$ is a regular
positive function for $0\leq\bar r\leq 1$, $-\infty <\bar z <\infty$,
$\bar r^2 + \bar z^2>0$.
The ADM-mass of $\tilde g(0)=G(0)^{4}g(0)$ is $m=3.877$, in particular it is
finite. For $\vert \bar z \vert$ large $G$ decays exponentially,
\begin{displaymath}
G(\bar r,\bar z;0) \propto \exp(-\alpha\vert\bar z\vert),\qquad \alpha=0.6856,
\end{displaymath}
it is thus zero at the physical origin $r=z=0$. There the Kretschmann
invariant diverges as $\tilde R_{abcd}\tilde R^{abcd}[\tilde g(0)]\propto
\exp(8\alpha\vert\bar z\vert)$, the physical metric thus has a curvature
singularity at the origin, where the minimal torus touches the axis of
symmetry.

\subsection{Existence of Solutions}

As stated above, the existence of a solution to the constraint, that is
a positive Green function for the elliptic operator $L_{\gamma}$ with boundary
condition given by Eq. (\ref{problem_conformal_2}), is equivalent to
positivity of the lowest eigenvalue $\lambda_{1}$ of $L_{\gamma}$.
Calling the eigenfunctions $\theta$ we obtain the eigenvalue problem
\begin{displaymath}
L_{\gamma}\theta = \lambda\theta, \qquad B_{\gamma}\theta\vert_{\bar r=1}=0,
\end{displaymath}
which is separable and the solutions can be written as
\begin{displaymath}
\theta_{\lambda m n}(\bar r,\varphi,\bar z)=\vartheta_{\lambda m n}
(\bar r)\sin(n\varphi + \alpha)\sin(m\gamma\bar z + \beta),
\end{displaymath}
where $\alpha$ and $\beta$ are arbitrary angles and we are left to solve an
equation for $\vartheta$,
\begin{displaymath}
\left(L_{\bar r\gamma}
      + \frac{m^2}{1-\gamma^{2}\bar r^{2}}
      + \frac{n^2}{\bar r^{2}}\right)\vartheta_{\lambda m n} =
                \lambda\vartheta_{\lambda m n},
\end{displaymath}
where $L_{\bar r\gamma}$ is the 'radial' part of the conformal Laplacian
$L_{\gamma}$,
\begin{equation}\label{L_reduced}
L_{\bar r\gamma}=-(1-\gamma^{2}\bar r^{2})\frac{\partial^2}
   {\partial \bar r^{2}}
      - \frac{1-3\gamma^{2}\bar r^{2}}{\bar r}\frac{\partial}{\partial \bar r}
      + \frac{3}{4}\gamma^{2}.
\end{equation}
Since the first eigenfunction is non-degenerate \cite{kazdan} and has
therefore to depend trivially on the angular coordinates $\varphi$ and
$\bar z$, we may set $m=n=0$ when looking for the lowest eigenvalue
$\lambda_{1}:$
\begin{equation}\label{existence-ODE}
L_{\bar r\gamma}\vartheta_{\lambda 00} =
             \lambda\vartheta_{\lambda 00}.
\end{equation}

The existence question has thus been reduced to an ODE of Sturm-Liouville
type which can be solved by standard methods. In this case routine
d02kef from the NAG Fortran Library \cite{NAG} was chosen to calculate
eigenvalues and eigenfunctions.

\begin{figure}[ht]
\begin{center}
\begin{psfrags}
 \psfrag{g}[]{$\gamma$}
 \psfrag{l1}[]{$\lambda_{1}$}
 \epsfxsize=3.375in\leavevmode\epsfbox{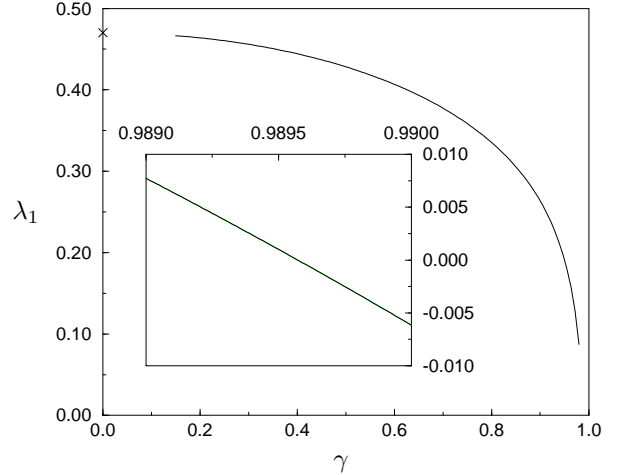}
\caption{The lowest eigenvalue $\lambda_1$ of the conformal Laplacian is
plotted as a function of $\gamma$, the insert shows that $\lambda_1$
gets negative for $\gamma>\gamma_{0}$. The value for $\gamma=0$ is marked
by an $\times$.}
\label{elam}
\end{psfrags}
\end{center}
\end{figure}

The lowest eigenvalue is found to be positive for $\gamma<\gamma_{0}=0.98957$,
and negative for $\gamma$ greater, see figure (\ref{elam}) and table (I).
For $\gamma=0$ Eq. (\ref{existence-ODE}) reduces to Bessel's equation:
\begin{displaymath}
\bar r^{2}\vartheta'' + \bar r\vartheta' + \lambda \bar r^{2}\vartheta =0,
 \qquad
\vartheta'+\frac{1}{4}\vartheta\vert_{\bar r=1}=0 .
\end{displaymath}
The regular solution is (apart from arbitrary scaling) $\vartheta=
J_0(\sqrt\lambda\bar r)$ ($J_{0}$ is the Bessel function of order $0$) where
$\lambda$ is chosen to satisfy the boundary condition
\begin{displaymath}
J_{0}'(\sqrt{\lambda}\bar r)+\frac{1}{4}J_{0}(\sqrt{\lambda}\bar r)
 \vert_{\bar r=1}=0 ,
\end{displaymath}
which yields $\lambda_{1}=0.4700$.

\subsection{Transforming the Constraint to ODEs}

Since the singular behavior of $G(\bar r, \bar z;\gamma)$ at $\Lambda$ makes
a direct numerical solution for the Green function inconvenient,
the first step in solving the system (\ref{problem_conformal_1},
\ref{problem_conformal_2}) is to regularize equation
(\ref{problem_conformal_1}). We split $G$ into the singular part $G_{0}$
and a regular part $\phi$ by
\begin{equation}\label{splitting}
G(\gamma) = G_{0}(\gamma) + \phi(\gamma).
\end{equation}
The asymptotic behavior of $\phi$ near $\Lambda$ is
\begin{displaymath}
\phi=\frac{m}{2\sqrt{2}} + O(G_{0}^{-1}),
\end{displaymath}
from the asymptotic behavior of the Green function (\ref{G_near_Lambda}),
$m$ is the ADM-mass of the physical metric $\tilde g$.
Calculation of the mass is thus as simple as evaluating the numerical
solution $\phi$ at the point $\Lambda$.
Insertion of (\ref{splitting}) into (\ref{problem_conformal_1},
\ref{problem_conformal_2}) yields an elliptic problem for $\phi$:
\begin{eqnarray*}
L_{\gamma}\phi & = & 0, \\
B_{\gamma}\phi\vert_{\bar r=1} & = &  -B_{\gamma} G_{0}\vert_{\bar r=1} \\
& = & \frac{G_{0}}{2}
\left.\left(\frac{\cos{\gamma\bar z}G_{0}^{2}}{\sqrt{1-\gamma^{2}}}-
\frac{1-2\gamma^{2}}{2(1-\gamma^{2})}\right)\right\vert_{\bar r=1} =:
   b(\bar z,\gamma).
\end{eqnarray*}
Note that the boundary condition becomes singular for $\gamma\rightarrow 1$.

The above partial differential equation in two dimensions can be turned into
a system of a countable set of uncoupled ODEs by decomposing $\phi$ and $b$
into a cosine series (the sine terms of the Fourier decomposition are zero
due to the symmetry with respect to reflection at the equator):
\begin{equation}\label{series_ansatz}
\phi=\sum_{n=0}^{\infty}\phi_{n}(\bar r)\cos(n\gamma\bar z),\qquad
b=\sum_{n=0}^{\infty} b_{n}(\bar r)\cos(n\gamma\bar z).
\end{equation}
This gives a boundary value ODE problem for every $n$ (prime denoting
differentiation with respect to $\bar r$):
\begin{eqnarray}\label{ODE}
\left(L_{\bar r\gamma} - \frac{n^{2}}{1-\gamma^{2}\bar r^{2}}\right)\phi_{n}
& = & 0 \\ \label{ODE_bc1}
\left.\left(\phi_{n}' + \frac{1-2\gamma^{2}}{4(1-\gamma^{2})}\phi_{n}
\right)\right\vert_{\bar r=1}          &=& b_{n}(\gamma),\\ \label{ODE_bc2}
\phi_{n}'\vert_{\bar r=0} &=& 0.
\end{eqnarray}
The last condition ensures regularity on $\bar r=0$, the $z-$axis.
By integrating the ODEs beyond $\bar r=1$ it is possible to continue the
solution smoothly inside the minimal tori to some finite extent,
but this will eventually lead to a nodal surface, and thus to a
curvature singularity.

Note that, for $n=0$, Eq. (\ref{ODE}) corresponds to Eq.
(\ref{existence-ODE}), and correspondingly only $\phi_{0}$ blows up for
$\gamma_{0}$, the higher order terms pass through $\lambda_{1}=0$ smoothly.

The equations are solved by a shooting and matching method (NAG routine
d02agf \cite{NAG}) for $n=1,\dots,N$, where $N$ is suitably chosen to make
the truncation error for $\varphi$ reasonably small.
\begin{figure}[ht]
\begin{center}
\begin{psfrags}
 \psfrag{g}[]{$\gamma$}
 \psfrag{m}[]{$m$}
 \epsfxsize=3.375in\leavevmode\epsfbox{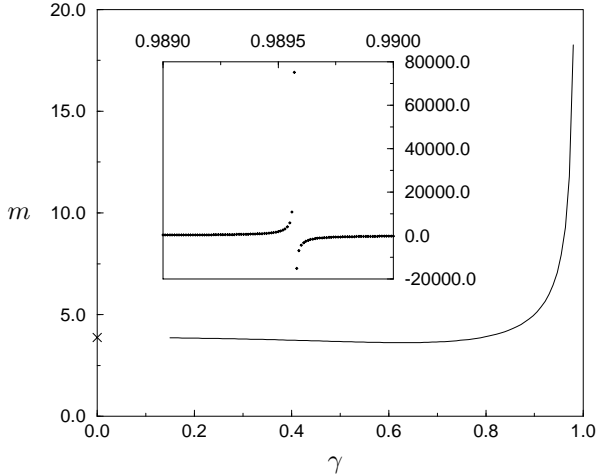}
\caption{The ADM mass is plotted as a function of $\gamma$, the insert shows
the pole of $m$ at $\gamma=\gamma_{0}$. The value for $\gamma=0$ is marked
by an $\times$.}
\label{m}
\end{psfrags}
\end{center}
\end{figure}

\subsection{Tests of Numerical Accuracy}

For every approximate solution $$\phi_{N}=\sum_{n=0}^{N}\phi_{n}(\bar r)
\cos{(n\gamma\bar z)}$$ with cutoff-parameter $N$ one may ask how well
the Hamiltonian constraint is satisfied. Due to numerical inaccuracies
and cutting off the Fourier series at finite $N$, Eqs.
(\ref{problem_conformal_01}) and
(\ref{problem_conformal_02}) will not be satisfied identically, but instead
there will appear two functions $E_{1}$ and $E_{2}$ such that
\begin{displaymath}
\bigtriangleup_{\delta}\psi = E_{1}\not\equiv 0,\qquad B_{\delta}\psi = E_{2}
\not\equiv 0
\end{displaymath}
where $E_{1}$, $E_{2}$ should be small in a sense that has yet to be
determined. From the Fourier series expansion it can be seen,
that increasing $N$ will increase $\vert E_{1}\vert$ too, since
the conformal Laplacian annihilates every single term in the above Fourier
series, $L_{g}\phi_{n}\cos{(n\gamma\bar z)}=0$, for $\phi_{n}$ known exactly.
$E_{1}$ is thus simply the sum of all numerical inaccuracies of the
individual Fourier
terms, and the sole purpose of increasing $N$ is to yield better satisfaction
of the boundary conditions, that is to decrease $E_{2}$.

Let us deal with the smallness of $E_{1}$ first. In the case of a vacuum
spacetime, where the scalar curvature has to vanish, there does not exist an
a priori measure for $E_{1}$. Following \cite{bernstein} the error can be
written as $\tilde R=16\pi\rho_{res}$, where $\rho_{res}$ can be interpreted
as the mass-energy density of matter for a non-vacuum spacetime.
The criterion of accuracy derived from this viewpoint is the smallness of
the total amount of mass represented by $\rho_{res}$ as compared to the
ADM-mass.

The residual of the Hamiltonian constraint is defined as
\begin{displaymath}
\rho_{res} = \frac{8}{16\pi}\psi^{-5} L_{\delta}\psi
\end{displaymath}
yielding the residual mass
\begin{displaymath}
m_{res} = \int_{V}\vert \rho_{res}\vert\,dV = \int_{V}\vert\rho_{res}\vert
\psi^{6}\,d^{3}x.
\end{displaymath}
Since the solution has been constructed in toroidal coordinates,
one gets better results, if derivatives are computed with respect to these,
so $m_{res}$ is rather evaluated in the form
\begin{displaymath}
m_{res} = \frac{1}{2\pi}\int_{V}\psi\Omega^{-1/2}\vert L_{g}G\vert \,
\bar r d\bar r d\bar z d\varphi.
\end{displaymath}
The volume of integration has to be chosen carefully. Since the domain of
computation is infinite with respect to the physical metric, integration
over the whole computational domain would yield an infinite result, the
integration was thus restricted to $\sqrt{r^{2}+z^{2}}\leq 50\,m$.

An alternative way to control the numerical error that is especially useful
to control fulfillment of the boundary conditions can be constructed by
putting the ADM mass into an explicitly positive form analogous to the
expression derived for Brill-waves in \cite{brill}. From Eq.
(\ref{problem_conformal_01}) we get ($\partial$ is the flat derivative
operator)
\begin{displaymath}
0=\frac{\bigtriangleup_{\delta}\psi}{\psi}=\partial^{2}\log\psi +
(\partial\log\psi)^{2}
\end{displaymath}
which can be integrated with respect to the flat background metric to give
\begin{displaymath}
\int_{\partial M}\frac{n^{a}\partial_{a}\psi}{\psi}\,dA +
\int_{M}\left(\frac{\partial\psi}{\psi}\right)^{2}\,dV = 0.
\end{displaymath}
The boundary $\partial M$ consists of spatial infinity and the boundary
torus $\Sigma$, inserting the definition of the ADM mass $m$ yields
\begin{displaymath}
m = \frac{1}{2\pi}\left(
\int_{\Sigma}\frac{n^{a}\partial_{a}\psi}{\psi}\,dA +
\int_{M}\left(\frac{\partial\psi}{\psi}\right)^{2}\,dV  \right).
\end{displaymath}
Using the boundary condition $n^{a}\partial_{a}\psi+\psi p/4 = 0$
allows to calculate the first integral explicitly as
\begin{displaymath}
m_{\Sigma}:=\frac{1}{2\pi}\int_{\Sigma}\frac{n^{a}\partial_{a}\psi}{\psi}\,dA
 =-\frac{1}{8\pi}\int_{\Sigma}p\,dA=\frac{\pi}{2}
\end{displaymath}
resulting in
\begin{equation}\label{mass_positive}
m = \frac{\pi}{2} + \frac{1}{2\pi}
\int_{M}\left(\frac{\partial\psi}{\psi}\right)^{2}\,dV=:m_{\Sigma}+m_{V}.
\end{equation}
The minimal mass of the metrics in the sequence considered is thus bounded
from below by $m>\pi/2$, the numerically calculated minimal value of the
mass is $3.623$ at $\gamma=0.628$.

In table (I) the relative errors of the numerically calculated expressions
$m_{\Sigma}$ and $m_{res}/m$ and the choices of $N$ are listed
along with other results for some values of $\gamma$.

As the coordinate range of $\bar z$ approaches the entire real line for
$\gamma\rightarrow 0$, the cosine series (\ref{series_ansatz})
has to be replaced by a cosine integral that can be written down explicitly
as is discussed in Appendix (\ref{expl_sol}), accordingly the cutoff
parameter $N$ of the Fourier series has to be increased for $\gamma$
decreasing to keep the accuracy approximately constant. For practical
reasons only few numerical computations were done for $\gamma<0.15$, so that
the region $0<\gamma<0.15$ is left out of the figures displayed.

\section{instability of the minimal tori}\label{instability}
%
An interesting question concerning
surfaces of minimal area is whether they are stable --  that is true minimal
surfaces. Stability of a minimal surface $\Sigma$ means
that for all possible one-parameter (normal) variations $\Sigma_{t}$ of the
surface $\Sigma=\Sigma_{0}$ the second derivative of the area satisfies
$A''(0)\geq 0$. A normal variation can be
written as $\phi n^{a}$ where $n^a$ is a smooth unit normal vector field along
$\Sigma_{0}$ and $\phi$ is a smooth function on $\Sigma$.
The second variation of the area is then given by
\begin{displaymath}
A''(0)= -\int_{\Sigma}\phi[\bigtriangleup_{\Sigma} \phi
+(\tilde R_{ab}n^{a}n^{b}+\tilde p_{ab}\tilde p^{ab})\phi]d\,A,
\end{displaymath}
where $\Sigma$ is now taken to be our minimal toroidal boundary surface
specified by $\bar r=1$ in the physical geometry,
$\bigtriangleup_{\Sigma}$ is the Laplacian on $\Sigma$, $\tilde R_{ab}$ the
Ricci curvature of $\tilde g$ on $\Sigma$, and $\tilde p_{ab}$ the extrinsic
curvature of $\Sigma$.
Stability holds iff all eigenvalues of the operator $-(\bigtriangleup_{\Sigma}
+(\tilde R_{ab}n^{a}n^{b}+\tilde p_{ab}\tilde p^{ab}))$
are nonnegative for a minimal surface.
If $\Sigma$ is of toroidal topology it is known \cite{galloway}
that it can only be stable if the minimal torus is totally geodesic,
which it cannot be in our case as was concluded from the form of $p_{ab}$
given in Eq. (\ref{pab}).
Nevertheless it is interesting to compute the spectrum (or at least the
first few eigenvalues, since this has to be done numerically).
Using the Gauss law and the constraint conditions $\tilde R=\tilde p=0$, and
denoting the scalar curvature of $\Sigma$ as ${^{2}\tilde R}$ the
eigenvalue problem becomes
\begin{displaymath}
-\bigtriangleup_{\Sigma}\phi + \frac{1}{2}({^{2}\tilde R}-\tilde p_{ab}
 \tilde p^{ab})\phi=\nu\phi.
\end{displaymath}
Straightforward computation yields
\begin{eqnarray*}
\bigtriangleup_{\Sigma} &=&
G^{-4}\left(\frac{1}{1-\gamma^{2}}\frac{\partial^{2}}{\partial\bar z^{2}}
+\frac{\partial^{2}}{\partial\varphi^{2}}\right), \\
{^{2}\tilde R} &=&  \frac{4G^{-4}}{1-\gamma^{2}}\left(
G^{-2}{G,_{\bar z}}^{2} - G^{-1}G,_{\bar z\bar z}\right), \\
\tilde p_{ab}\tilde p^{ab} &=& G^{-4}\left(p_{ab}p^{ab} - \frac{p^{2}}{2}
\right)= \frac{1}{2}G^{-4}(1-\gamma^{-2})^{-1}.
\end{eqnarray*}
With some rearrangement of terms the eigenvalue equation becomes
\begin{displaymath}
-\frac{\partial^{2}\phi}{\partial\bar z^{2}} -(1-\gamma^{2})
\frac{\partial^{2}\phi}{\partial\varphi^{2}} + V\phi
=\nu G^{4}(1-\gamma^{2})\phi,
\end{displaymath}
\begin{displaymath}
V=-\frac{1}{4} + 2\left(G^{-2}{G,_{\bar z}}^{2}-G^{-1}G,_{\bar z\bar z}\right),
\end{displaymath}
which has to be solved with periodic boundary conditions in $\bar z$ and
$\varphi$.
Since the geometry is axially symmetric, so is $G$, and the equation
separates. The eigenfunctions can be written as
\begin{displaymath}
\phi(\bar z,\varphi)=u_{n\nu}(\bar z)v_{n}(\varphi),
\end{displaymath}
where $v_{n}(\varphi)=\sin{n(\varphi+\alpha)}$ are the axial eigenfunctions
degenerate in the axial phase shift $\alpha$ and for every $n=0,1,2,\dots$
one gets an equation for the $u_{n\nu}(\bar z)$:
\begin{equation}\label{stab_eq}
- u_{n\nu}'' + \left(
(1-\gamma^{2})n^{2}+V\right)u_{n\nu}=\nu_{n} G^{4}(1-\gamma^{2})u_{n\nu},
\end{equation}
which is a regular self-adjoint Sturm-Liouville problem with periodic boundary
conditions for $\gamma\neq 0,\gamma_{0}$. The spectrum is thus purely discrete
and bounded below (see e.g. Ref. \cite{levitan}, Sec. 1.4).

As a result of using periodic boundary conditions the eigenfunctions are
grouped in pairs by their number of zeros (see e.g. Theorem 1.4.2 of
Ref. \cite{levitan}):
The eigenfunctions $u_{k}$, k=0,1,2\dots, with eigenvalue $\nu_{k}$
have $k$ zeros for $k$ even, and
$k+1$ zeros for $k$ odd, that is $u_{1}$ has no zero, $u_{2}$ and $u_{3}$
both have two zeros, $u_{4}$ and $u_{5}$ four, etc., where the endpoints of
the interval are identified.

\begin{figure}[ht]
\begin{center}
\begin{psfrags}
 \psfrag{gamma}[]{$\gamma$}
 \psfrag{n}[]{$\nu$}
 \epsfxsize=3.375in\leavevmode\epsfbox{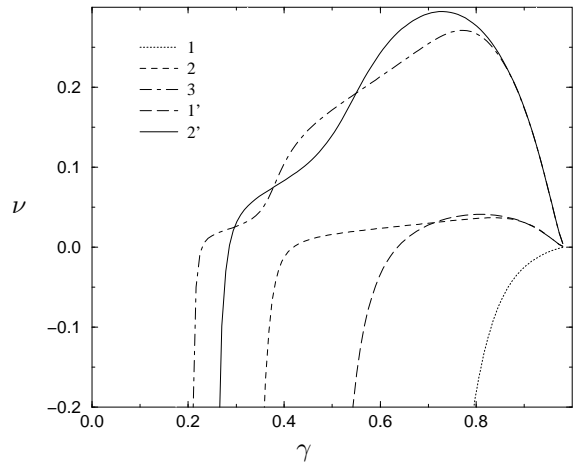}
\caption{The first five eigenvalues $\nu_{0l}$ ($n=0$)
are plotted as functions of $\gamma$ from top down ($l = 1,\dots, 5$).
The curves labelled $1,2,3$ are the first three branches of eigenvalues with
symmetric eigenfunctions, while $1',2'$ label the first two branches of
eigenvalues with antisymmetric eigenfunctions.}
\label{m0}
\end{psfrags}
\end{center}
\end{figure}

\begin{figure}[ht]
\begin{center}
\begin{psfrags}
 \psfrag{g}[]{$\gamma$}
 \psfrag{n}[]{$\nu$}
 \epsfxsize=3.375in\leavevmode\epsfbox{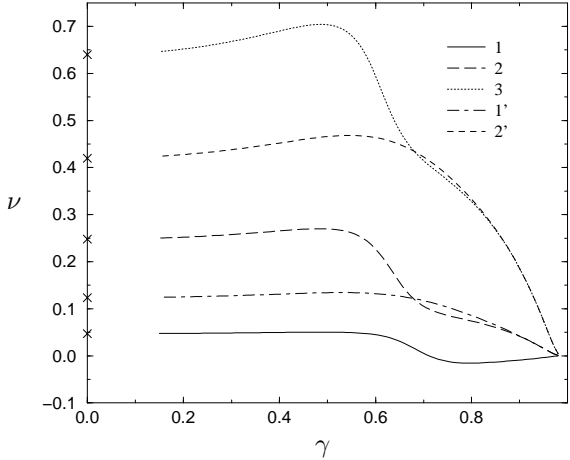}
\caption{The first five eigenvalues $\nu_{1l}$ ($n=1$) are
 plotted as functions of $\gamma$ from top down ($l = 1,\dots, 5$), the
 values for $\gamma=0$ are marked by $\times$'s. The curves labelled $1,2,3$
 are the first three branches of eigenvalues with symmetric eigenfunctions,
 while $1',2'$ label the first two branches of eigenvalues with antisymmetric
 eigenfunctions.}
\label{m1}
\end{psfrags}
\end{center}
\end{figure}

For practical purposes it is convenient to convert the periodic boundary
conditions to unmixed boundary conditions, which is made possible by
noting that for Eq.(\ref{stab_eq}) it is sufficient to consider only
eigenfunctions that are  with respect to $\bar
z=0$. If a periodic solution $u$ exists that is neither
symmetric nor antisymmetric,
then the whole solution space is spanned by the linear combinations
$u(x)\pm u(-x)$ and the general solution is periodic, since with $u(x)$ being
a periodic solution, so is $u(-x)$. The eigenvalue is of multiplicity two in
these cases, which show up in figures (\ref{m0},\ref{m1}) as the
crossing points of branches associated with symmetric and antisymmetric
eigenfunctions that have both the same number of zeros.

The problem of finding
solutions with periodic boundary conditions can thus be split into the
easier problems of finding all periodic symmetric and antisymmetric solutions,
equivalent to imposing the unmixed boundary conditions
\begin{eqnarray*}
u'(0) &=& u'\left(\frac{\pi}{\gamma}\right)=0
 \quad\mbox{for symmetric solutions},\\
u(0) &=& u\left(\frac{\pi}{\gamma}\right)=0
  \quad\mbox{for antisymmetric solutions},
\end{eqnarray*}
so that standard methods for Sturm-Liouville problems can be applied.
Here routine d02kef from the NAG Fortran Library \cite{NAG} was chosen to
calculate eigenvalues and eigenfunctions, as was done already for Eq.
(\ref{existence-ODE}).

The results are given in figures (\ref{m0},\ref{m1}) for $n=0,1$. For all
values of $\gamma$ one can find at least one negative eigenvalue,
demonstrating that the tori are not stable, and are thus
only saddle points of the area functional. In particular the lowest
eigenvalue for $n=0$ is found to be negative for all $\gamma$, and all
other eigenvalues for $n=0$ become negative as well for $\gamma$ small
enough, see Fig. (\ref {m0}). For $n=1$ only the lowest eigenvalue is negative
for $\gamma>0.71$, see Fig. (\ref {m1}). For $n=2$ the situation is similar:
again only the lowest eigenvalue is negative, but in the much smaller region
$\gamma>0.97$. For $n>2$ the situation is not totally clear:
if negative eigenvalues appear, they have to occur very close to
$\gamma_{0}$.
By explicit computation of the area of neighboring tori with
$\bar r=const.$, it was found that the minimal tori
have minimal area for $\gamma<0.891$, and maximal area for $\gamma>0.891$
among neighboring tori of constant $\bar r$.

In the limiting case $\gamma=\gamma_{0}$ the Green function $G$ diverges,
and all eigenvalues $\nu$ go to zero. In the other limiting case,
$\gamma = 0$, the coordinate range is the entire real line, and the
Sturm-Liouville problem becomes singular. The potential $V$ falls off
exponentially to the value $n^{2}-1/4$ for large
$\vert\bar z\vert$ -- note that
this value is negative for $n=0$ and positive for $n>0$. While nothing
special happens to the part of the spectrum with $n>0$, see Fig. (\ref{m1})
for $n=1$, the situation is different for $n=0$. The general solution $u$ to
Eq. (\ref{stab_eq}) for $\gamma=n=0$ and arbitrary $\nu$ is asymptotically
trigonometric (see e.g. Ref. \cite{courant}, section 5.11), that is
\begin{displaymath}
u=\alpha\sin(\bar z + \beta) + O(1/\bar z ),
\end{displaymath}
which is oscillatory with finite amplitude near $\bar z=\pm\infty$.
Periodic boundary conditions thus can not be imposed on these solutions,
for small $\gamma$ all eigenvalues decrease unboundedly, marking a growing
instability of the minimal surface with respect to axially symmetric
deformations.
\section{Apparent Horizons}\label{ah}

As was shown before, the minimal tori considered here are all unstable,
so that they locally can be deformed to an outer trapped surface
\cite{galloway}. By asymptotic flatness the region of compact outer trapped
surfaces has to be finite, the outer boundary is called apparent horizon,
which is a stable marginally outer trapped surface
(see Ref. \cite{galloway} and references cited therein).
Noting that $\tilde g$ is analytic by construction of the conformal factor as
the solution of Eqs. (\ref{ODE}-\ref{ODE_bc2}) and not flat (by analyticity
it would
have to be flat everywhere if it were flat in some region, but this can not
be, since the mass is positive by Eq. (\ref{mass_positive})), it can be
concluded from  Ref. \cite{galloway} that the apparent horizon has spherical
topology.

The existence and properties of apparent horizons have received some
interest in the context of the cosmic censorship hypothesis
\cite{pen-cimento}, where they play a double role as in Penrose's singularity
theorem \cite{penrose} and also as a local (albeit only weak) analogue
of the global concept of an event horizon. The
second is practically exploited in the apparent horizon boundary condition
\cite{ahbc} used in numerical relativity and also was used in a new approach
for the definition of a black hole itself \cite{hayward}.

\begin{figure}[ht]
\begin{center}
\begin{psfrags}
 \psfrag{g}[]{$\gamma$}
 \psfrag{A}[]{$\frac{A}{16\pi m^{2}}$}
 \epsfxsize=3.375in\leavevmode\epsfbox{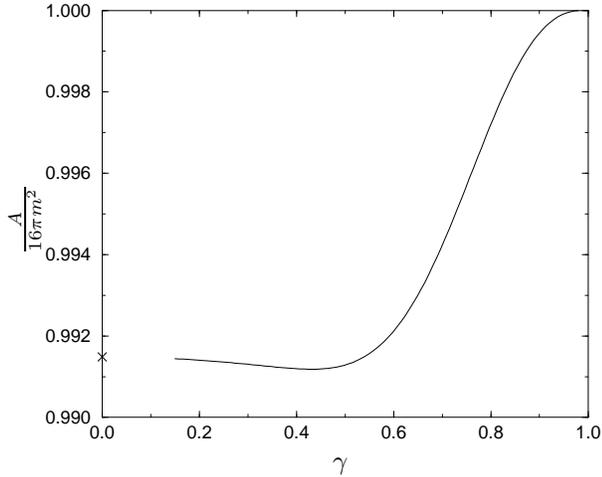}
\caption{Check of the Penrose isoperimetric inequality, the value for
 $\gamma=0$ is marked by $\times$.}
\label{pen}
\end{psfrags}
\end{center}
\end{figure}

The apparent horizons were located numerically and analyzed with respect to
the Penrose and hoop conjectures. Penrose \cite{pen-cimento} formulated
the conjecture that if a collapsing body satisfies the inequality
\begin{displaymath}
m<\sqrt{\frac{A}{16\pi}},
\end{displaymath}
where $m$ is the ADM mass and $A$ the area of a trapped surface, then cosmic
censorship should be violated. As pointed out by Horowitz \cite{horo}, if $A$
is the minimal area that surrounds a trapped surface (in the time
symmetric case identical to the apparent horizon) and satisfies the
above inequality, then cosmic censorship does not hold.
For the apparent horizons found here the inequality is always
violated, the Penrose conjecture thus confirmed, see figure (\ref{pen}).

\begin{figure}[ht]
\begin{center}
\begin{psfrags}
 \psfrag{g}[]{$\gamma$}
 \psfrag{c}[]{}
 \psfrag{ce}[l][tr]{${C_{e}}/{4\pi m}$}
 \psfrag{cp}[l][br]{${C_{p}}/{4\pi m}$}
 \epsfxsize=3.375in\leavevmode\epsfbox{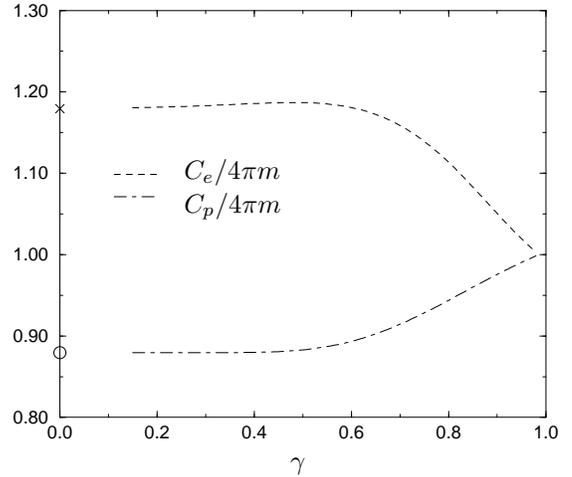}
\caption{Equatorial and polar hoop quotients, the values for
 $\gamma=0$ are marked by $\times$ and $\circ$.}
\label{hoop}
\end{psfrags}
\end{center}
\end{figure}

The hoop conjecture, due to Thorne \cite{thorne}, states that an
apparent horizon will exist if and only if a mass $m$ gets compacted into
a region such that the circumference $C$ satisfies $C\leq 4\pi
m$ in every direction. The exact definition of what should be meant
by $m$ and $C$ is left open. For a precise definition of $m$ one
would need a notion of quasilocal mass which is available only
for the spherically symmetric case (no gravitational waves), here
the ADM mass was chosen.
Also for a general definition of the circumference various definitions
would be possible. In an axially symmetric geometry which is also symmetric
with respect to reflection at the equator, straightforward definitions of
circumference exist along the equator ($C_{e}$) and along a meridian
circle ($C_{p}$):
\begin{eqnarray*}
C_{e} &=& \left. 2\pi\psi^{2}R\right\vert_{\theta=\pi/2},\\
C_{p} &=& 4\int_{0}^{\pi/2}d\theta\,\psi^{2}\sqrt{R'^{2}+R^{2}},
\end{eqnarray*}
where $R$ is the radial coordinate in standard polar coordinates
$(R,\theta,\varphi)$. The results for $\frac{C}{4\pi m}$
are given in Fig. (\ref{hoop}) and table (I).

For the problem of locating apparent horizons in the axially symmetric
case a variety of different methods is available by now. If one considers
configurations which are symmetric (or almost symmetric) with respect to
reflection at the equator, one can reasonably hope that the apparent horizon
will be expressible as a function $R(\theta)$. This gives the ODE
\begin{displaymath}
\frac{R''-\frac{R'^{2}}{R}}{1+\frac{R'^{2}}{R^{2}}}
- 4 R^{2}\frac{\psi_{R}}{\psi}
+R'\left(4\frac{\psi_{\theta}}{\psi} + \cot\theta\right)
- 2 R=0,
\end{displaymath}
which has to be solved with the boundary conditions
\begin{displaymath}
R'(0)=R'(\pi/2)=0.
\end{displaymath}
This is done numerically using a shooting and matching
method (NAG \cite{NAG} routines d02agf and d02ebf were used in two different
codes for horizon finding).

The results are shown in figures (\ref{pen}), (\ref{hoop}) and
table (I). For
$\gamma\rightarrow \gamma_{0}$ the apparent horizon becomes spherical and
pinches off
to infinity as has been discussed for the analogous case when the Yamabe
number goes to zero for a manifold without boundary in Ref. \cite{beig+om}.
For $\gamma<\gamma_{0}$ the apparent horizon is oblate,
but not very strongly.

\section{discussion}\label{discussion}

In a previous paper \cite{tid} the notion of toroidal conformal symmetry
was introduced to label a class of initial data where a $U(1)\times U(1)$
conformal isometry is present. Working on a compactified background metric
allows to easily exploit this conformal symmetry to decompose the
Lichnerowicz conformal factor in a double Fourier series on the group
orbits and give the solution in terms of a countable family of uncoupled
ODEs on the orbit space.
The present paper applies these methods to construct and analyze initial
data containing a marginally outer trapped torus.
Special emphasis is given to
the  investigation of the limiting cases when the sequence of solutions
intersects the boundary of conformal superspace.

To get a better understanding of the results it seems useful to look at
the usually studied case of vacuum asymptotically flat data without
boundary, where the cases understood best are the time-symmetric, notably
the axially symmetric subclass known as Brill-waves \cite{om-brill}.

In order to simplify this comparison I will first discuss generalizations
of the data considered in this paper.
The obvious generalization is to drop conformal flatness and consider
deformations of the base metric $g$ of the form
\begin{equation}\label{q-deform}
g = e^{2Aq(\bar r)}\left(\frac{{d\bar r}^{2}}{1-\gamma^{2}\bar r^{2}} +
      (1-\gamma^{2}\bar r^{2}){d\bar z}^{2}\right) + \bar r^{2}{d\varphi}^{2},
\end{equation}
which is the analogue of the Brill-wave ansatz \cite{brill} while keeping the
conformal symmetry, and is currently investigated. The general axially
symmetric case corresponds to form
functions $q(\bar r,\bar z)$ that depend on both $\bar r$ and $\bar z$.
Within the present ansatz one could consider certain non-axially symmetric
situations by placing $\Lambda$ off the axes (the sets of
fixed points of the isometries), the physical metric then has no isometries
whatsoever -- since an isometry of $g$
carries over to $\tilde g$ only if $\Lambda$ is a fixed point of this isometry
\cite{CKV}.

The usually studied case is recovered, finally, by imposing regularity on all
of $R^{3}$ instead of the $\tilde p=0 $ boundary condition, this case has
been discussed in \cite{tid}, the publication of numerical results is in
preparation.

For these traditionally considered data sequences of metrics which approach
the boundary of the space of initial data have been studied both analytically
and numerically and tested for apparent horizons.
Two complementary paths have been taken, an overview is given in
\cite{om-brill}.
Sequences of metrics for which the Yamabe number goes to zero
have been studied analytically by Beig and \'O Murchadha \cite{beig+om}, who
proved that asymptotically the mass diverges in indirect proportion to the
Yamabe number and that an apparent horizon would always appear before the
mass diverges.
Abrahams et al. \cite{ahst} on the other hand studied families of
metrics of fixed mass with a singular limit for which no
apparent horizon appeared even in the limiting case.

The limiting case $\gamma\rightarrow\gamma_{0}$ of the present paper is
similar to the one
where the Yamabe number goes to zero. As the Sobolev quotient defined in
Eq. (\ref{sobolevQ}) is just an extension of the Yamabe number including an
additional boundary term, the mathematical and physical situation is
analogous. As the mass diverges, the apparent horizon becomes more and more
spherical and pinches off to infinity. Going beyond that critical point
the mass becomes infinitely negative. Whereas in the
no-boundary case the amplitude parameter $A$ of Eq. (\ref{q-deform}) can be
increased without bound, which gives rise to an infinite number of analogous
singularities in the mass function (to be discussed in forthcoming work), here
the parameter $\gamma$ is
bounded from above by $\gamma\leq 1$, and the mass appears to stay negative
for $\gamma>\gamma_{0}$. A proof that the Sobolev quotient will become
negative was unsuccessful due to the finite range of $\gamma$: In the
Brill-wave case an
estimate can be obtained with simple test-functions by just increasing the
amplitude enough \cite{om-brill} -- here the same trick can not be carried
out. $\gamma_{0}$ comes in fact very close to the upper boundary
$\gamma=1$,
and so a very good choice of test function would have to be made.

The other case where the sequence of solutions hits the boundary of
conformal superspace, $\gamma\rightarrow 0$, is fundamentally different.
While in the first case the base geometry stays regular, the situation when
the boundary torus touches the axis can be considered as singular,
nevertheless the mass and location of the apparent horizon remain finite,
but a curvature singularity develops at the origin.

Table (I) lists some numerical results for different values
of $\gamma$.
\begin{table}
\begin{center}
\begin{tabular}{c|c|c|c|c|c|c}\label{only_table}
$\gamma$ & $m$ & $\lambda_{1}$ & $C_{p}/C_{e}$  & $m_{res}/m$  &
$\frac{m_{\Sigma}}{\pi/2}-1$ & $N_{max}$ \\
\hline
0.0 & 3.8771 & 0.4700   & 0.4758 & -  & -  & - \\
\hline
0.2 & 3.8442 & 0.4639 & 0.7447 & $3.7\times 10^{-10}$ & $ -3.7\times
10^{-6}$ &
100 \\
\hline
0.4 & 3.7444 & 0.4443 & 0.7421 & $1.6\times 10^{-9}$ & $6.4\times 10^{-7}$
  & 37 \\
\hline
0.5 & 3.6771 & 0.4283 & 0.7440 & $2.1\times 10^{-9}$ & $ -7.0\times 10^{-9}$
  & 33 \\
\hline
0.6 & 3.6272 & 0.4068 & 0.7567 & $4.1\times 10^{-9}$ & $-3.3\times 10^{-8}$
  & 24 \\
\hline
0.7 & 3.6571 & 0.3775 & 0.7891 & $2.5\times 10^{-8}$ & $-3.9\times 10^{-8}$
  & 19 \\
\hline
0.8 & 3.9173 & 0.3353 & 0.8476 & $7.6\times 10^{-8}$ & $-1.1\times 10^{-7}$
  & 15 \\
\hline
0.9 & 5.0129 & 0.2636 & 0.9287 & $2.5\times 10^{-7}$ & $-1.1\times 10^{-7}$
  & 12 \\
\hline
0.989 & 226.13 & 0.0773 & 0.9998 & $4.8\times 10^{-4}$ & $7.8\times 10^{-7}$
  & 7\\
\hline
\end{tabular}
\caption{Results from numerical calculations, numbers are rounded to the
last digit.}
\end{center}
\end{table}

The technique of using conformal symmetry together with compactification
had several technical advantages.
First of all, the ODEs (\ref{ODE}) and (\ref{existence-ODE}) resulting from
the Hamiltonian constraint and existence question can be handled
with relatively small numerical expertise and desktop workstations.
The use of the NAG numerical Fortran library \cite{NAG} yields an
excellent documentation of all numerical routines that have been
used.

Another point in favor of the method is that it works well for strong data --
that is near $\gamma=\gamma_{0}$ in our sequence, only the $n=0$ term in the
Fourier series blows up when the mass does.

If there are additional asymptotic regions (representing black holes),
discrete symmetries result from an arrangement compatible to the conformal
isometry as discussed in \cite{tid}.

It is hoped, that the technical approach taken here and in \cite{tid},
namely to use compactification and conformal Killing vector fields may also
prove useful in other situations.
\acknowledgements
The author thanks R. Beig for his support, encouragement and for
many stimulating discussions.
This work was financially supported by Fonds zur F\"orderung der
wissenschaftlichen Forschung, Project No. P09376-PHY.

\begin{appendix}

\section{The explicit limiting solution}\label{expl_sol}

In the limiting situation $\gamma=0$ the constraint equation
(\ref{problem_g0_1}) can be
solved explicitly. Making the ansatz
\begin{displaymath}
G = \frac{\sqrt{2}}{\sqrt{\bar r^{2} + \bar z^{2}}}+\phi
\end{displaymath}
(analogous to (\ref{splitting})),
we get
\begin{eqnarray*}
\bigtriangleup \phi & = & 0,\\
\left.\left(\frac{\partial \phi}{\partial\bar r}+\frac{1}{4}\phi\right)
\right\vert_{\bar r=1} & = &
 (1+\bar z^{2})^{-3/2} - \frac{1}{4}(1+\bar z^{2})^{-1/2}
\end{eqnarray*}
from Eqs. (\ref{problem_g0_1}) and (\ref{problem_g0_2}).

We solve by cosine transformation:
\begin{displaymath}
\phi(\bar r,\bar z)=
\frac{2}{\pi}\int_{0}^{\infty}\cos({\omega\bar z})f_{\omega}(\bar r)\,d\,
\omega,
\end{displaymath}
resulting in a family of Bessel equations for regular functions
$f_{\omega}(\bar r)$ parameterized by $\omega$:
\begin{displaymath}
f_{\omega}''+\frac{1}{r}f'_{\omega}-\omega^{2}f_{\omega}=0.
\end{displaymath}
The regular solutions are
\begin{displaymath}
f_{\omega}(\bar r)=c(\omega)I_{0}(\omega \bar r),
\end{displaymath}
where $I_{n}$ and $K_{n}$ used below are modified Bessel functions of
order $n$ and
$c(\omega)$ is determined from the cosine transformed boundary condition
\begin{eqnarray*}
f'_{\omega}(\bar r)+\frac{1}{4}f_{\omega}(\bar r)
& = & {\sqrt\frac{2}{\pi}}\int_{0}^{\infty}\cos({\omega\bar
z}) \\
&\times&\left( (1+\bar z^{2})^{-3/2} - \frac{1}{4}(1+\bar z^{2})^{-1/2}\right)
\,d\omega\\
& = & \frac{1}{4}{\sqrt\frac{2}{\pi}}\left(4\omega K_{1}(\omega)-
K_{0}(\omega)\right).
\end{eqnarray*}
 as
\begin{displaymath}
c(\omega)= {\sqrt\frac{2}{\pi}}
\frac{4\omega K_{1}(\omega)-K_{0}(\omega)}
{4\omega I_{1}(\omega) + I_{0}(\omega)}.
\end{displaymath}
We get an explicit formula for the Green function,
\begin{eqnarray}\label{G_explicit}
G(\bar r,\bar z)  &=& \frac{\sqrt{2}}{\sqrt{\bar r^{2} + \bar z^{2}}}
+\phi = \frac{\sqrt 2}{\sqrt{\bar r^{2} +\bar z^{2}}} \nonumber\\
& + & \frac{2}{\pi}\int_{0}^{\infty}\cos({\omega\bar z})
\frac{4\omega(K_{1}(\omega)-K_{0}(\omega)}
{4\omega I_{1}(\omega) + I_{0}(\omega)} I_{0}(\omega\bar r)\,d\omega.
\end{eqnarray}
$\phi$ is a regular function of $\bar r$, $\bar z$, the integrand
diverges only logarithmically for small $\omega$ and exhibits exponential
falloff for large $\vert\bar z\vert$. Its value at the point at infinity
gives the ADM mass $m=3.877$ by $m=2\sqrt{2}\phi(\Lambda)$.

Alternatively the Green function can be written as
\begin{equation}\label{G_together}
\begin{array}{rcl}
G(\bar r,\bar z ) & = & \frac{2}{\pi}\int_{0}^{\infty}\cos({\omega\bar z})
\frac{p(\omega, \bar r)}{q(\omega)}\,d\omega,\\
p(\omega, \bar r) & = & 4\omega(I_{1}(\omega)K_{0}(\omega\bar r)+
K_{1}(\omega)I_{0}(\omega\bar r))\\
& &+K_{0}(\omega\bar r)I_{0}(\omega)
-I_{0}(\omega\bar r)K_{0}(\omega),\\
q(\omega) & = & 4\omega I_{1}(\omega) + I_{0}(\omega).
\end{array}
\end{equation}
using the identity \cite{gradshteyn}
\begin{displaymath}
\frac{1}{\sqrt{\bar r^{2} + \bar z^{2}}} =
 \frac{2}{\pi}\int_{0}^{\infty}\cos({\omega\bar z})
K_{0}(\omega \bar r)\,d\,\omega.
\end{displaymath}

Positivity of $G(\bar r,\bar z)$ was checked by numerical evaluation,
the behavior of $G$ for large $\vert \bar z \vert$ can be seen analytically by
integration of (\ref{G_together}) in the complex plane.
Both $p$ and $q$ are positive and symmetric on the real line,
but $q$ has infinitely many zeros for purely imaginary arguments,
leading to corresponding poles of $p/q$.
Observing, that the factor $\exp{(-\mbox{Im}(\omega)\bar z)}$
makes contributions to the integral small for large $\bar z$ in the upper
complex half plane (for $\bar z\rightarrow -\infty$
the lower half respectively)
yields an asymptotic formula for the integral by choosing a path that
encircles only the first pole (an asymptotic series results from summing
over all poles),
\begin{displaymath}
G(\bar r,\bar z ) = c(\bar r)\exp{(-\alpha\vert\bar z\vert)},
\end{displaymath}
where $\alpha=0.6856$ is the first zero of $q(\omega)$ on the imaginary axis
and
\begin{displaymath}
c(\bar r) = 4\,i\,\mbox{Residual}\left(\frac{p(\omega,\bar r)}{q(\omega)},
\omega=0.6856\right).
\end{displaymath}
\end{appendix}
%
%

%

\begin{references}
\bibitem{pos_mass}
For an overview on the positive mass theorem of Schoen and Yau \cite{sy}
and its relation to the Yamabe problem see Ref. \cite{lp}.
\end{references}
\end{document}